\documentclass[twocolumn,showpacs,amsmath,amssymb]{revtex4}
\usepackage{graphicx}
\begin{document}

\title{Ferroelectricity of Ice Nanotubes inside Carbon Nanotubes}
\author{Chuanfu Luo}
% \altaffiliation[also at]{Physics Department, Nanjing University.}
% \email{luochuanfu@gmail.com}
\author{Wei Fa}
\author{Jinming Dong}
 \email{jdong@nju.edu.cn}
 \affiliation{
   Group of Computational Condensed Matter Physics, National Laboratory of Solid State
   Microstructures and Department of Physics, Nanjing University, Nanjing, 210093, People's Republic of China
 }
\date{\today}

\begin{abstract}
We report that ice nanotubes with odd number of side faces inside
carbon nanotubes exhibit spontaneous electric polarization along
its axis direction by using molecular dynamics simulations. The
mechanism of this nanoscale quasi-one-dimensional ferroelectricity
is due to low dimensional confinement and the orientational order
of hydrogen bonds. These ferroelectric fiber structural materials
are different from traditional perovskite structural bulk
materials.
\end{abstract}
\pacs{77.84.-s, 85.35.Kt, 61.46.Fg, 68.08.De} \maketitle

%\section{\label{sec:intro}introduction}

Whether ferroelectric ice exists is a question that has long
fascinated
researchers\cite{bernal_jacs33,pauling_jacs35,slater_jcp41,jackson_jcp95}.
Although one water molecule has a significant dipole moment,
normal hexagonal ice I\emph{h} does not possess ferroelectricity
and it is very difficult to get an ice phase even with a weak
whole polarization. Several year ago, the ferroelectricity of ice
films grown on ultra-clean platinum was detected, although only a
small proportion (0.2\%) of the molecules are
aligned\cite{su_prl98,bramwell_nat99}. Most recently, Fukazawa
\emph{et al.} have succeeded in making Ice XI in the laboratory
and suggests existence of ferroelectric ice in the
universe\cite{fukazawa_06}.

Confinement of matter on the nanometer scale can induce phase
transitions not seen in bulk systems \cite{Gelb_rpp99,Hung_apl05}.
In biology as well as in natural and synthetic materials, water is
often tucked away in tiny crevices inside proteins or in porous
materials. Therefore, water confined in nanoscale
quasi-one-dimensional (Q1D) channels is of great interest to
biology, geology, and materials science. Recent investigations
have increased our understanding of confined water, showing that,
in nanoscopic proportions, many water properties differ
drastically from those of bulk water
\cite{koga_nat00,levinger_sci02,choe_prl05}. In particular, an
excellent model is water confined in carbon nanotubes, which have
been adopted in many previous studies
\cite{koga_jcp00,koga_nat01,koga_pya02,bai_jcp03,noon_cpl02,
hummer_nat01,werder_nano01,mashl_nano03,
mann_prl03,naguib_nano04,sriraman_prl05,striolo_jcp05,
mamontov_jcp06,striolo_nano06}.
%{dellago_prl03,waghe_jcp02,liu_jcp05,moulin_prb05,liu_prb05,striolo_jcp06,hanasaki_jcp06,}
Among these interesting investigations, Koga \emph{et al.}
investigated water inside single-walled carbon nanotubes (SWCNTs)
of diameter 1.1$\sim$1.4 nm and found that water forms ice
nanotubes composed by a rolled square ice sheet
\cite{koga_nat01,koga_pya02,bai_jcp03}. Recent neutron scattering
studies, combined with molecular dynamics (MD) simulations,
revealed that water in SWCNTs of diameter 1.4 nm forms a
core-shell structure, in which the square ice sheet described by
Koga \emph{et al.} is coupled with a chainlike configuration at
the center of the shell \cite{kolesenikov_prl04,reiter_arxiv06}.

Recently, investigations of nanoscale ferroelectric materials are
very active due to its potential application
\cite{spanier_nano06,bune_nat98,ahn_sci04}. Because of its
particular structures, the ice nanotubes inside SWCNTs may exhibit
ferroelectricity.

In this paper, we report that ice nanotubes with odd number of
side faces inside SWCNTs exhibit spontaneous electric polarization
along its axis direction by using MD simulations.  The origin of
ferroelectricity of these ice nanotubes is due to the strong
direction preferred hydrogen-bonds between water molecules and the
Q1D confinement inside SWCNTs, which is different from traditional
ferroelectric bulk materials, such as displacive perovskite-based
oxides.

%\section{\label{sec:model}model and computational method}
The water-water interaction is modelled by the TIP4P model
\cite{jorgensen_jcp83}, which has been used in many simulations
and given theoretical results in well consistent with experiments
\cite{matsumoto_nat02,koga_jcp00,koga_nat00,koga_nat01,koga_pya02}.
The Ewald summation is used to deal with the long range Coulomb
interaction and the Lennard-Jones interaction between water
molecules is cut off at 9.0 \AA. The interactions between water
molecules and SWCNTs are contributed only by the Lennard-Jones
potential between the oxygen and carbon atoms\cite{hummer_nat01},
which can be written as
\begin{eqnarray}{
U_{oc}= 4\epsilon_{oc} \sum_{i=1}^{N_o} \sum_{j=1}^{N_c}
[(\frac{\sigma_{oc}}{r_{ij}})^{12}-
(\frac{\sigma_{oc}}{r_{ij}})^6]},
\end{eqnarray}
where $N_o$ is the total number of oxygen atoms, $N_c$ is the
total number of carbon atoms, and $r_{ij}$ is the distance between
the oxygen and carbon atoms. The parameters are taken as
$\epsilon_{oc}=0.1029$ kcal mol$^{-1}$ and $\sigma_{oc}=3.280$
\AA, obtained from Lorentz-Berthlot combining rules of
$\epsilon_{oc}=\sqrt{\epsilon_{oo} \times \epsilon_{cc}}$ and
$\sigma_{oc}=(\sigma_{oo}+\sigma_{cc})/2$. The parameters,
$\epsilon_{oo}=0.1549$ kcal mol$^{-1}$ and $\sigma_{oo}=3.154$
\AA\ are taken from the TIP4P model, $\epsilon_{cc}=0.06835$ kcal
mol$^{-1}$ and $\sigma_{cc}=3.407$ \AA\ are taken from
multiple-shell fullerences and multiwalled carbon
nanotubes\cite{popov_prb02,lu_prb94}. To speed up MD simulations,
we assumed that the carbon atoms on SWCNTs are distributed
continuously, which is well known as the continuum model and has
been used in a lot of systems \cite{tabar_prep04}. Thus the
potential between one oxygen atom and a SWCNT is written as
\begin{eqnarray}
{ U(r)= n_d \int u(x)d\Sigma },
\end{eqnarray}
where $r$ and $x$ represent the distances of the oxygen atom to
the axis of SWCNT and the surface element $d\Sigma$, respectively.
$n_d$ is the mean surface density of atoms on SWCNT
\cite{girifalco_prb00,wu_prb05}. The C-C bond length of SWCNTs is
assumed to be 1.42 \AA. We checked the difference between the
formula (1) and (2) is within 2\% in our simulation conditions.

\begin{figure}
\includegraphics[]{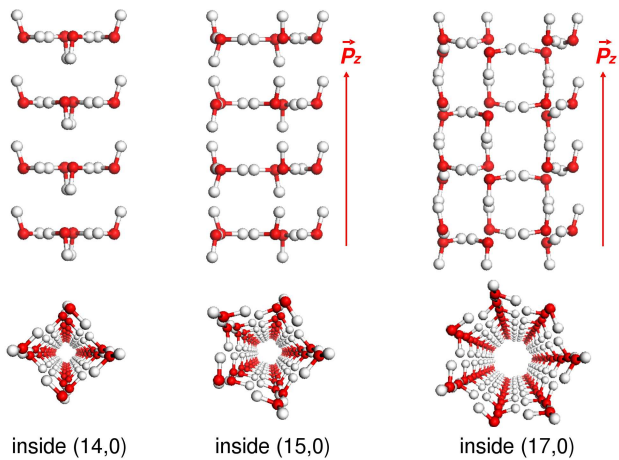}
\caption{\label{fig:Fig.1}(color online)Relaxed structures of ice
nanotubes inside SWCNTs. The red and gray balls denote oxygen and
hydrogen atoms, respectively. The ice nanotubes inside (15,0) and
(17,0) SWCNTs have spontaneous electric polarizations along its
axis direction. The polarization vectors are illustrated by
$\vec{P}_z$ and arrows.}
\end{figure}

The MD simulations are performed by a modified TINKER package
\cite{tinker_ref}, in which the velocity-Verlet algorithm for
integration of equation of motion and the Berendsen thermostat
algorithm are used. The time step is set to be 0.5 fs. The motion
of translation and rotation of the whole water molecules are
eliminated during the MD simulations to calculate temperature. The
simulation box is taken as a 50\AA$\times$50\AA$\times$$L$ cubic
box, where $L$ is the length along the axis of SWCNT. The periodic
boundary conditions along the axis are used. Then, the real
axial-pressure ($P_{zz}$) is the original calculated value times
2500 $\mathrm{\AA}^2/S$, where $S$ is the area of the cross
section of SWCNT.

%%TIP4P/Ice\cite{abascal_jcp05}  TIP4P/2005\cite{abascal_jcp05_2}

\begin{figure}
\includegraphics[]{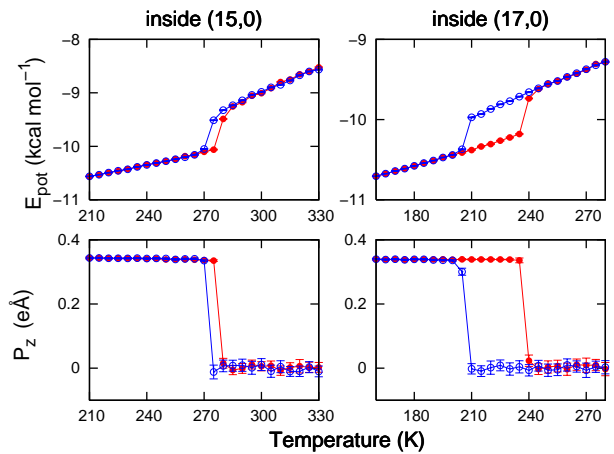}
\caption{\label{fig:Fig.2}(color online)The potential energy per
water molecule ($\mathrm{E_{pot}}$, in which the water-SWCNT
interaction energy is excluded) and polarization per layer
($\mathrm{P_{z}}$) versus temperature. The left and right panels
are of the ice nanotubes inside (15,0) and (17,0) SWCNTs,
respectively. The red filled circles and blue unfilled circles
indicate the heating and cooling processes, respectively. The
statistic standard errors are shown by the error bars, and lines
are to guide the eye.}
\end{figure}

Following Koga \emph{et al.}
\cite{koga_jcp00,koga_nat01,koga_pya02}, we first simulated water
inside SWCNTs at constant temperature ($T$) and $P_{zz}$ by using
NPT assemble. We chose a high axial-pressure ($P_{zz}$=200MPa)
condition and four types of zigzag SWCNTs, which are (14,0),
(15,0), (16,0), and (17,0) with diameters of 10.96 \AA, 11.74 \AA,
12.53 \AA, and 13.31 \AA, respectively. During our MD simulations,
the temperature was lowered stepwise staring from 450K to 100K and
then heated stepwise back. The MD simulation time at each
temperature is 2ns (10ns or more near the phase transition point).
Some relaxed structures of the ice nanotubes and the polarized
directions are both shown in Fig.1. These results consistent with
those of Koga \emph{et
al.}\cite{koga_jcp00,koga_nat01,koga_pya02,bai_jcp03}. As
illustrated in Fig.1, the ice nanotubes have 4, 5, and 7 side
faces corresponding to inside (14,0), (15,0), and (17,0) SWCNTs,
respectively. The ice nanotube inside (16,0) has 6 side faces but
not shown in Fig.1. From the viewpoint of application, we think
that water or ice will probably be enclosed inside SWCNTs to
produce Q1D devices. So, our later MD simulations were performed
at constant $T$ and volume ($V$ or $L$), known as NVT assemble.
The initial densities are taken by considering the results at high
$P_{zz}$. In our simulations, the $L$ is taken as 109.2 \AA\ and
the total number of water molecules in the simulation box is
$40\times n$, where $n$ is the number of side faces and refers 4,
5, 6 and 7 corresponding to the four types of ice nanotubes inside
(14,0), (15,0), (16,0), and (17,0) SWCNTs. The relaxed structures
of the ice nanotubes and the polarized direction are almost the
same as those at high $P_{zz}$ shown in Fig.1. It is found that
the ice nanotubes with odd number of side faces are ferroelectric,
while those with even number of side faces are antiferroelectric.
The polarizations of the relaxed ice nanotubes with $n$=5 and 7
are 0.351 e\AA\ and 0.347 e\AA\ per layer of water molecules. If
we take a bundle of these ferroelectric ``fibers'' arranged in a
dense packed hexagonal 2D lattice with a interspace distance of
3.4 \AA\ (the same distance as that in graphite), the polarized
rates are 0.15 C m$^{-2}$ and 0.12 C m$^{-2}$, respectively (the
rate of BaTiO$_3$ is 0.26 C m$^{-2}$).

\begin{figure}
\includegraphics[]{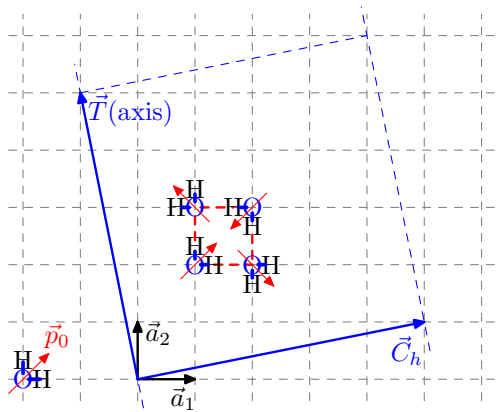}
\caption{\label{fig:Fig.3}(color online)Sketch of an unrolled 2D
ice sheet, in which, the basis vectors ($\vec{a}_1$ and
$\vec{a}_2$), the chiral vector ($\vec{C}_h$), and the axis vector
($\vec{T}$) are shown. $\vec{C}_h=n\vec{a}_1+m\vec{a}_2$
corresponds to a $\langle n,m \rangle$ ice nanotube. The
``Bernal-Cowler-Pauling bulk ice rule'' is that every water
molecule serves as a double donor and a double acceptor of
hydrogen bonds, and every water molecule is hydrogen-boned to
exactly four nearest-neighbor
molecules\cite{bernal_jacs33,pauling_jacs35}. In the 2D ice sheet
and ice nanotube, the ``ice rule'' is still satisfied and the
water network looks like a chessboard consisting of many square
lattices. The square hydrogen-bond network is denoted in dashed
lines and the details of the hydrogen bonds are shown at the
center by four water molecules. The red arrow at a site denotes
the dipole moment of a water molecule and is to direct its
orientation. The example ice nanotube in this figure is $\langle
5,1 \rangle$. }
\end{figure}

The potential energy ($\mathrm{E_{pot}}$) and spontaneous electric
polarization along axis ($\vec{P}_{z}$) versus $T$ of two
ferroelectric ice nanotubes are plotted in Fig.2. It is found that
the ferroelectric-paraelectric of the ice nanotubes strongly
depends on the solid-liquid phase transition. The ferroelectric
phase transition behavior of the ice nanotubes is much different
from traditional ferroelectric bulk materials such as BaTiO$_3$,
whose transition occurs at solid state phase and obeys the
Curie-Weiss law. As an order parameter, the $\mathrm{P_{z}}$ of an
ice nanotube drops at the solid-liquid phase transition point
consisting with its $\mathrm{E_{pot}}$-$T$ curve. This
ferroelectric-paraelectric phase transition is a first-order
transition coupled with violent structural change. The
ferroelectricity is due to the special structures of ice nanotubes
and the orientations of water molecules, which details are
discussed as follows.

\begin{figure}
\includegraphics[]{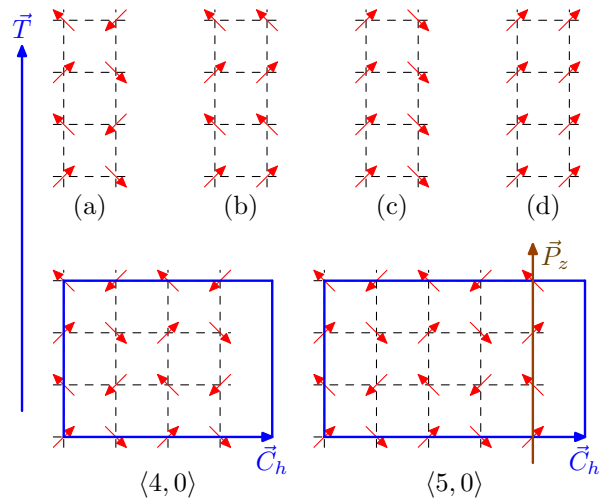}
\caption{\label{fig:Fig.4}(color online)The upper panel: four
types of orientations of water molecules consisting one side face
in a $\langle n,0 \rangle$ ice nanotube; The lower panel: unrolled
2D hydrogen-bond network of two example ice nanotubes, $\langle
4,0 \rangle$ and $\langle 5,0 \rangle$. The red arrows indicate
the orientations of water molecules as show in Fig.3. The
potential energies of (a), (b), (c) and (d) are in increasing
order, so the type (a) is preferred. A $\langle 4,0 \rangle$ ice
nanotube has 4 water molecules per layer with 2 upwards and 2
downwards, so it has no polarization at every layer. Every layer
of a $\langle 5,0 \rangle$ ice nanotube has 5 water molecules with
3 upwards and 2 downwards and has a polarization along the axis
due to the one redundant upward water molecule. All the layers
have the same directions of polarization along the axis since the
``ice rule'' forbids any opposite one. The ``ice rule''
accumulates the polarizations at every layer and generates a whole
spontaneous electric polarization $\vec{P}_z$ along the axis. The
blue boxes are to denote the unit cell.}
\end{figure}

Just like for SWCNTs, we can analyze the structure of an ice
nanotube by rolling a 2D ice sheet. For simplicity, we use a
right-angle water molecular model to describe the 2D ice sheet. As
illustrated in Fig.3, a 2D ice sheet consists of many square
lattices and an ice nanotube can be described by a pair of numbers
$\langle n,m \rangle$. By using this description, the three ice
nanotubes shown in Fig.1 can be denoted as $\langle 4,0 \rangle$,
$\langle 5,0 \rangle$, and $\langle 7,0 \rangle$, respectively. A
pair of $\langle n,m \rangle$ can describe the main frame of water
network of an ice nanotube but fail to describe the orientations
of water molecules. If we use the dipole moment of a water
molecule $\vec{p}_0$ (0.45 e\AA\ in the TIP4P model and 0.38 e\AA\
in experiment) to direct its orientation, we can get another
equivalent description of the ``ice rule'' in a 2D ice sheet and
so in an ice nanotube: the projections of all diploe moments
aligned along the basis vectors ($\vec{a}_1$ and $\vec{a}_2$ in
Fig.3) must be in the same direction. That is to say, there are
only two types of orientation for two nearest water molecules,
parallel or perpendicularly.

To understand why the water molecules prefer the orientations
shown in Fig.1, a very simple model is used to compare the
energies of different isomers. We take the approximation:
$En\simeq E_{LJ}+ E_{HY}+ E_P$, here $E_{LJ}$ is the Lennard-Jones
potential, $E_{HY}$ is the potential owned by hydrogen bonds, and
the $E_P$ is the potential due to dipole-dipole interaction which
can be written as
\begin{eqnarray}{
 E_P= \frac{1}{4\pi\epsilon_0} \sum_{ij} \frac{\vec{p}_i \cdot \vec{p}_j - 3(\hat{r}_{ij}
\cdot \vec{p}_i)(\hat{r}_{ij} \cdot \vec{p}_j)}
{{|\vec{r}_{ij}|}^3}},\end{eqnarray} where $\epsilon_0$ is the
permittivity of space, $\vec{p}_i$ and $\vec{p}_j$ are the dipole
moments of water i and j, respectively. $\vec{r}_{ij}$ is the
displacement from water i to j, and $\hat{r}_{ij}$ is its unit
vector. For the first order approximation, the summation is over
all nearest-neighbor sites. Although it might overestimate $E_P$,
it can give us a clearer qualitative physical picture. The
$E_{LJ}$ and $E_{HY}$ terms lead to the ``ice rule'' and decide
the main frame of hydrogen-bond network. The $E_P$ term due to
dipole moments is much smaller than the other two terms but has an
effect on the orientation of water molecules. At the same basis
vectors (assuming $|\vec{a}_1|=|\vec{a}_2|=r_0$), different
isomers with different orientations of water molecules have the
same $E_{LJ}$ and $E_{HY}$ but different $E_P$, which makes us
compare the $E_P$ term only. A pair of perpendicular water
molecules are $p^2_0/4\pi\epsilon_0 r^3_0$ lower in energy than a
pair of parallel ones, so the orientations of two nearest water
molecules prefer the perpendicular type.

Now, we discuss why only the ice nanotubes with odd number of side
faces possess ferroelectricity. We first classify a local water
network to four types if we just consider the nearest pairs of
water molecules. The upper panel of Fig.4 shows an example of the
side faces of a $\langle n,0 \rangle$ ice nanotube. The type (a)
has the lowest energy without polarization and is preferred when
an ice nanotube is formed. The type (b) is polarized along the
axis although in a small higher energy. The potential energy
differences between the four types are very small. As an example
of $\langle 4,0 \rangle$ ice nanotube, the type (b) is only
$4.8\times10^{-2}$ kcal mol$^{-1}$ ($2.1\times10^{-3}$ eV) per
water molecule higher in energy than type (a). The lower panel of
Fig.4 illustrates why a $\langle 5,0 \rangle$ ice nanotube is
ferroelectric while a $\langle 4,0 \rangle$ ice nanotube is
antiferroelectric. The reason is that a $\langle 5,0 \rangle$ ice
nanotube has one ferroelectric side face of type (b) while a
$\langle 4,0 \rangle$ ice nanotube has only antiferroelectric side
faces of type(a). This consideration can be easily extended to
other ice nanotubes. The origin of this type of ferroelectricity
is the Q1D confinement and the orientational order of hydrogen
bonds. We get the conclusion that, an ice nanotube with odd number
of side faces has at lease one redundant polarized water molecule
at every layer and the ``ice rule'' accumulates them along the
axis and generates a whole spontaneous electric polarization.

It is worthy to mention that, during our MD simulations, many fast
quenched ice nanotubes exhibit ferroelectricity whatever odd or
even number of side faces, which lie in small higher energies than
those shown above. The side faces of these isomers are mainly type
(b) and very stable at ice phase. So, if we apply an external
electric field and cool the water fast, we might get ice nanotubes
with larger polarization than those shown in this paper. It is
also found that some $\langle 5,0 \rangle$ and $\langle 7,0
\rangle$ ice nanotubes with defects are slightly helix and
ferroelectric too. In experiment, the orientations of water
molecules can be detected by neutron scattering or infrared
spectra, and the ferroelectricity can be reflected by second-order
nonlinear optical coefficients.

%\section{\label{sec:summary}summary}
In summary, ice nanotubes inside SWCNTs with odd number of side
faces can exhibit ferroelectricity. These nanoscale Q1D
ferroelectric fiber structural materials are different from
traditional ferroelectric bulk materials both in
ferroelectric-paraelectric phase transition behavior and the
original mechanism. They may be applied in many fields, such as
high sensitive senors, nanoscale electric machines, nonvolatile
memory devices, and so on. In addition, the similar type of
ferroelectric mechanism probably exists in other nanoscale porous
materials.

%\section{\label{sec:ackn}acknowledgements}
The authors thank Professor Ponder of Washington University for
his kind offering of the TINKER package. This work was supported
by the Natural Science Foundation of China under Grants No. 10474035
and No. A040108, and also from a Grant for State Key
Program of China through Grant No. 2004CB619004.

\end{document}